# Si/SiGe multi-channel superlattice structure epitaxial growth with segmented temperature control for Next-Generation Logic Devices

**Wenlong Yao[1], Zhigang Li[1,a], Guobin Bai[1], Jianfeng Gao[1,a], Jiahan Yu[1], Junfeng Li[1], Xiaolei Wang[1], Jun Luo[1]**

[1]Institute of Microelectronics of The Chinese Academy of Sciences, Beijing, 100029, China

**Abstract:** Stacking multiple Si/SiGe channels in advanced logic devices faces severe thermal budget accumulation, which degrades interfaces via Ge–Si interdiffusion and strain relaxation. Here we grew 2 + 2, 3 + 3 and 4 + 4 channel superlattices by reduced-pressure chemical vapor deposition and implemented a segmented temperature strategy: 650 °C for initial $Si_{0.8}Ge_{0.2}$ layers and 600 °C for subsequent layers. This strategy lowers the Ge diffusion coefficient to 5.6–7% of its value at 650 °C (Arrhenius estimate), suppressing interdiffusion and preserving pseudomorphic strain. The 4 + 4 channel stack exhibits clear XRD satellite peaks, fully coherent strain state (reciprocal space mapping), sharp interfaces (1.5–2.6 nm transition width) and low RMS roughness (0.08 nm). Quantitative analysis from bottom to top reveals that prolonged high-temperature exposure broadens bottom interfaces and dilutes Ge concentration (from ~20 % to ~18.5 %), while the top stack maintains design targets. This work provides a process–physics understanding of thermal budget effects in multi-channel superlattices and establishes a high-quality material foundation for advanced logic devices beyond 2 nm node.

[a]**Corresponding authors.**
E-mail: gaojianfeng@ime.ac.cn, lizhigang2@ime.ac.cn

## 1. Introduction

Continued scaling of logic devices beyond the 2 nm node requires three-dimensional architectures such as gate-all-around FETs (GAAFETs) and complementary field-effect transistor (CFET)[1–5]. By vertically stacking n- and p-type devices, the CFET effectively reduces the standard cell area while maintaining the drive current [6–11] (**Figure 1**). A key enabler is the Si/SiGe superlattice stack: Si layers serve as conductive channels, while SiGe sacrificial layers are later replaced by dielectric isolation [12–14]. Recent studies on Si/SiGe superlattice for CFET applications have focused on validating the feasibility of superlattice structures and stacking multiple channel layers. Intel and IMEC demonstrated SiGe stacked superlattice structures for CFETs in 2020 and 2025, respectively, directly confirming the feasibility and necessity of multi-channel SiGe superlattice structures [15–17]. Huang et al. [15] realized a 4 + 4 channel superlattice structure constructed in a self-aligned manner. Loo et al. [16] employed a relatively high growth temperature to develop 1 + 1 channel and 3 + 3 channel structures for CFET and preliminarily investigated the interface quality of the 1 + 1 channel structure. Meanwhile, Loo et al. [17] highlighted the challenges faced by SiGe superlattice structures for CFETs, including preventing relaxation and the formation of threading dislocations, ensuring sufficient Ge content for the etch selectivity between SiGe and Si, and maintaining superlattice quality while achieving high growth rates. As channel counts increase, however, the bottom layers experience prolonged high-temperature exposure, leading to thermal budget accumulation – a phenomenon that aggravates Ge–Si interdiffusion, broadens interfaces, and promotes strain relaxation [2, 18-21].

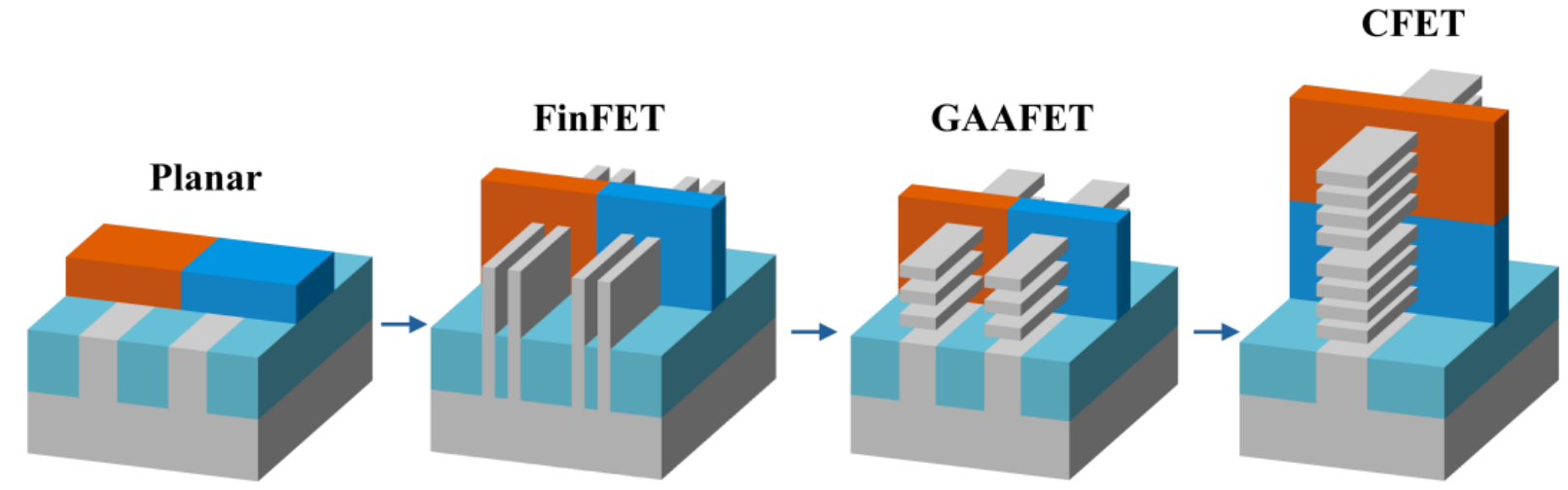

**Figure 1.** Transistor architecture evolution from planar transistors, to FinFETs, gate-all-around FETs (GAAFETs) and complementary field-effect transistors (CFETs).

In this work, we systematically investigate 2 + 2, 3 + 3 and 4 + 4 channel $Si/Si_{0.8}Ge_{0.2}$ | $Si_{0.6}Ge_{0.4}$ | $Si/Si_{0.8}Ge_{0.2}$ superlattices grown by reduced-pressure chemical vapor deposition (RPCVD). We demonstrate that a segmented strategy (650 °C – 600 °C) enables a fully coherent 4 + 4 stack with total thickness > 210 nm, sharp interfaces, and excellent periodicity. Moreover, we provide the first quantitative bottom-to-top analysis of interface thickness and Ge concentration, directly linking thermal budget accumulation to interfacial degradation. Our results not only solve a practical epitaxy challenge for logic devices but also offer insight into diffusion-limited interface engineering in Group-IV heterostructures.

## 2. Experimental Section

The stacked superlattice structures were grown on 200 nm Si (100) wafers using RPCVD. Three experimental groups were designed to study the influence of channel number on structural quality: 2 + 2, 3 + 3, and 4 + 4 channel superlattices. The high-Ge-concentration SiGe layer thickness was set to 50 nm for the 2 + 2 channel structure and 30 nm for the 3 + 3 and 4 + 4 channel structures. All other epitaxial process conditions were kept uniform, with only the number of stacking periods being varied. Prior to epitaxial layer deposition, an in-situ high-temperature pre-bake at 1120 °C for 120 s was performed to remove the native oxide layer. Epitaxial growth was carried out using hydrogen as the carrier gas with a flow rate of 22 slm and a chamber pressure of 80 Torr. Conventional precursor gases ($SiH_4$, DCS, and $GeH_4$ diluted in $H_2$) were used to grow the Si/SiGe superlattice structures containing two different Ge concentrations.

To address thermal budget accumulation in multi-channel stacks, a segmented temperature strategy was implemented. The $Si_{0.6}Ge_{0.4}$ layer and preceding layers were grown at 650 °C, while subsequent Si and SiGe layers were grown at 600 °C. This strategy mitigates interdiffusion, preserves interface sharpness, and maintains full strain, overcoming limitations of conventional isothermal growth, which can increase diffusion and reduce crystal quality. At these temperatures, Si growth rates decrease more than SiGe; therefore, silane was used as the Si precursor to maintain relatively high growth rates despite the lower temperature [15, 22–24]. A schematic and TEM cross-section of the four-channel structure are shown in **Figure 2**. Furthermore, to validate the segmented temperature control strategy, a comparative study was conducted on 4 + 4 channel superlattices grown at 600 °C, 650 °C, and segmented 650–600 °C.

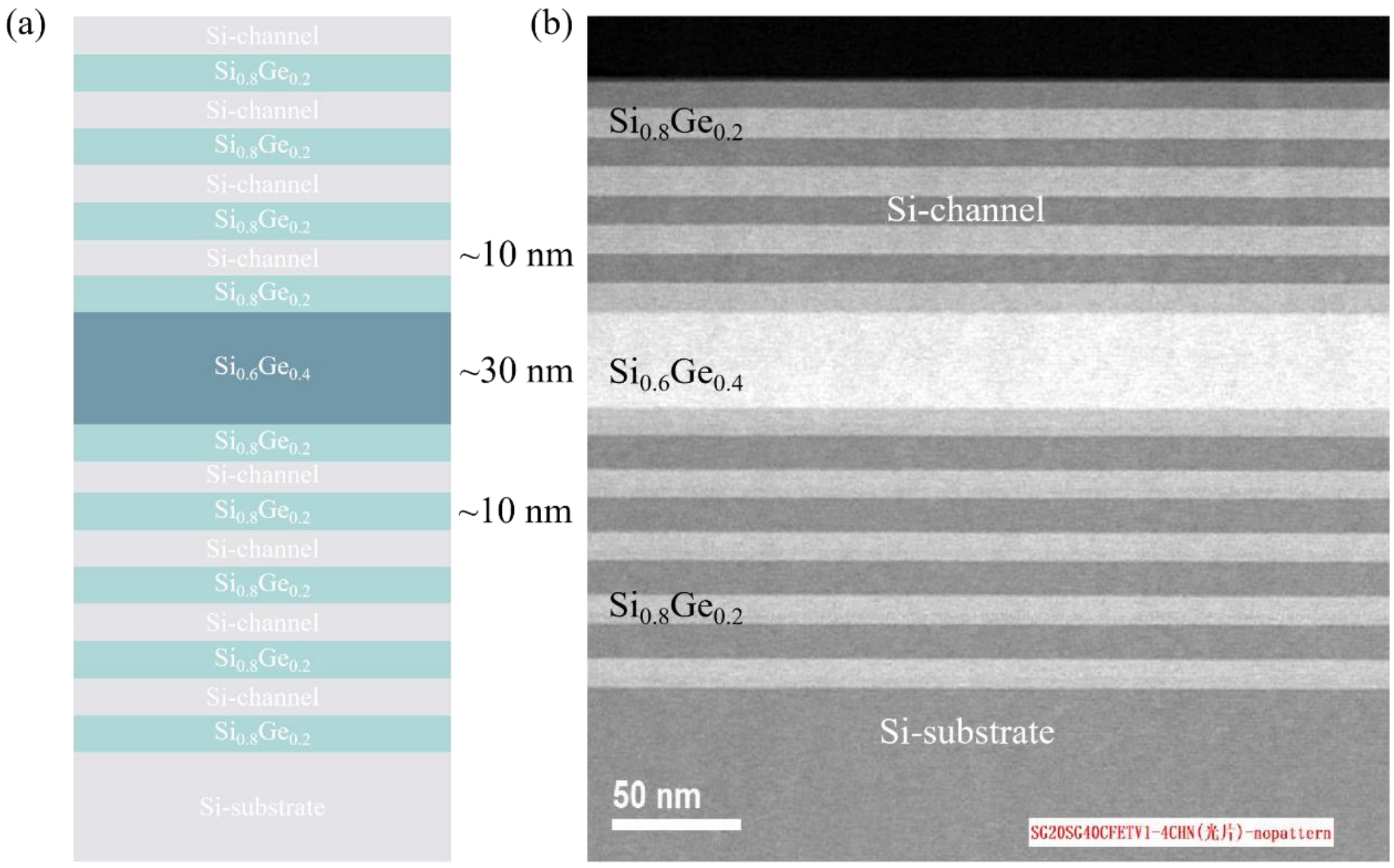

**Figure 2.** (a) Schematic cross-section and (b) cross sectional HAADF STEM image of a Si/SiGe multi-stack used for CFET devices with 4 + 4 channels for both the p-type and the n-type devices.

The quality of the superlattice structures was characterized using multiple complementary techniques. High-resolution X-ray diffraction (HRXRD) and reciprocal space mapping (RSM) were employed to measure structural properties, including strain quality, SiGe layer thickness, and Ge concentration. High-resolution transmission electron microscopy (HRTEM) was used to assess interface quality and layer thickness, while energy-dispersive X-ray spectroscopy (EDS) analyzed interface abruptness and Ge distribution. Atomic force microscopy (AFM) was used to characterize the surface roughness and morphology of the films. The combination of these multiple characterization techniques provides complementary information [25-26].

## 3. RESULTS and discussion

**Theoretical Basis of the Segmented Temperature Control Strategy**

The segmented temperature control strategy proposed in this work is grounded in two fundamental principles: diffusion kinetics and thermodynamics.

(i) The bottom layers in multi-channel stacks experience prolonged exposure to high temperatures, which significantly increases the risk of atomic interdiffusion. Compared with $Si_{0.8}Ge_{0.2}$, the $Si_{0.6}Ge_{0.4}$ layers contain higher Ge concentration. The intrinsically weaker Ge–Ge bond strength in these layers reduces the activation energy for interfacial diffusion and degrades their thermal stability, making them more susceptible to interlayer atomic diffusion.

According to the Arrhenius law of diffusion, the atomic diffusion coefficient exhibits an exponential relationship with temperature [27]:

$$D = D_0 \exp(-\frac{E_a}{kT}) \quad (1)$$

Based on the activation energies for Ge diffusion in SiGe reported in Ref. [27] (approximately 3.7 eV to 4.0 eV), reducing the growth temperature from 650 °C to 600 °C is estimated to decrease the diffusion coefficient of Ge atoms at the interface by approximately 1.16–1.25 orders of magnitude. This means that the diffusion coefficient is suppressed to about 5.6–7% of its original value, thereby effectively inhibiting interdiffusion of Ge atoms in the SiGe layers and maintaining sharp interfaces.

(ii) Critical thickness depends on growth temperature [28–29]. Higher temperatures provide energy for dislocation nucleation and glide, causing the initially metastable strained layers to gradually relax and reducing the effective critical thickness for maintaining pseudomorphic growth in the stack. Therefore, introducing a lower growth temperature at 600 °C during the later stages of multi-layer stacking preserves the interfaces already formed and increases the effective critical thickness of subsequent layers. This enables the 4 + 4 channel structure, with a total thickness exceeding 210 nm, to maintain a fully coherent strain state without misfit dislocations.

### Validation of the Segmented Temperature Control Strategy

To validate the effectiveness of the segmented temperature control strategy, a comparative experiment was conducted. 4 + 4 channel superlattice samples were fabricated under identical conditions (including thickness, composition, and precursors), but using three different temperature profiles: 600 °C, 650 °C, and segmented 650–600 °C. And the crystal quality, strain state, and surface morphology were systematically analyzed.

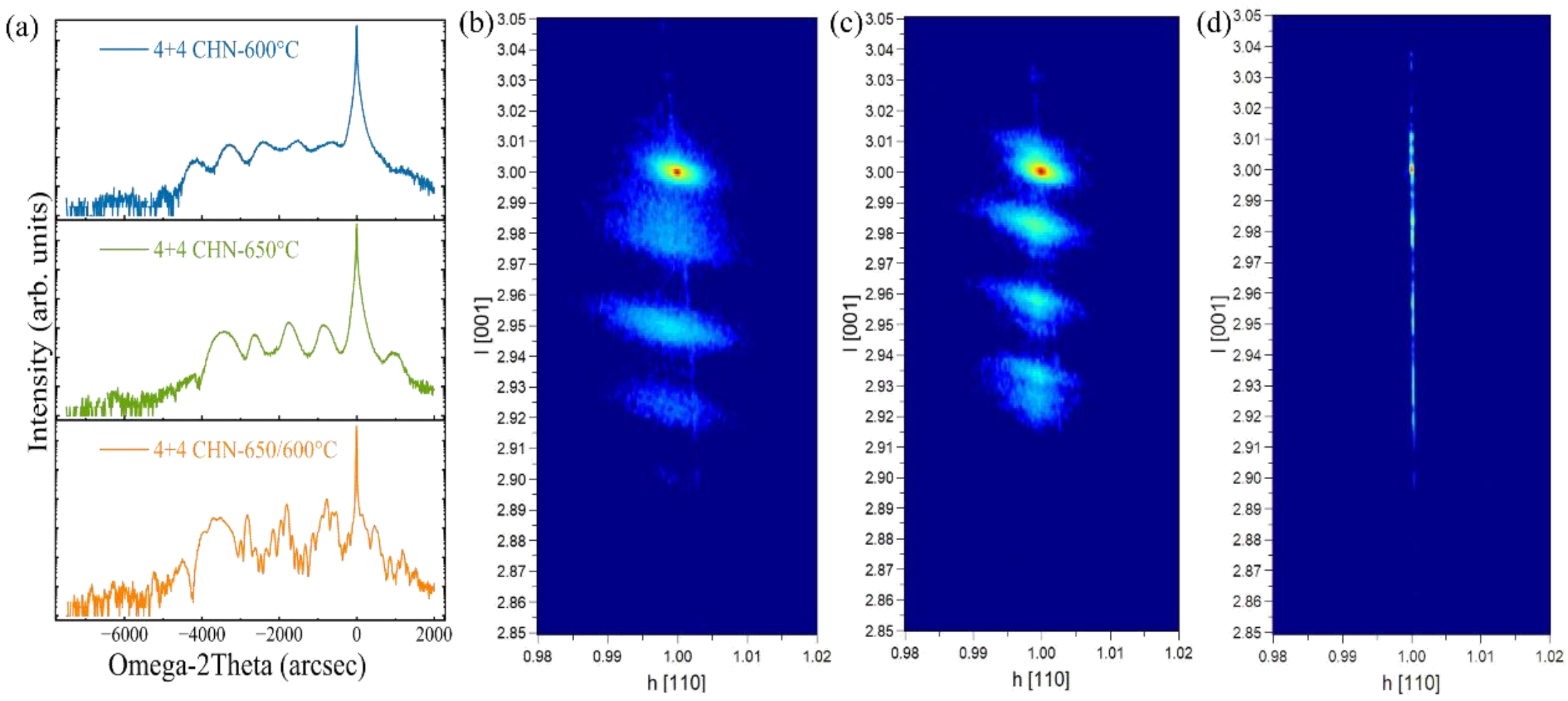

**Figure 3.** (a) HRXRD ω-2θ scans around the Si (004) Bragg reflection for the 4 + 4 samples grown at 600 °C, 650 °C, and segmented 650–600 °C. (b–d) Corresponding RSM around the Si (113) Bragg reflection.

As shown in **Figure 3(a)**, the constant 600 °C sample and constant 650 °C sample both exhibit the main satellite peaks, while these peaks are weak in intensity and shifted towards the Si substrate peak, and no higher-order diffraction peaks are observed. This indicates that severe interdiffusion has occurred in the two constant-temperature samples, leading to degradation of superlattice periodicity and a significant reduction in the Ge concentration of the $Si_{0.6}Ge_{0.4}$ layer. In contrast, the segmented temperature-controlled sample exhibits clear multi-order satellite peaks, demonstrating its excellent periodicity and atomically abrupt interfaces.

This quality difference is further confirmed in the asymmetric (113) RSM analysis. For the two constant-temperature samples, diffraction spots show significant broadening and dispersion, proving that interfacial interdiffusion and misfit dislocation proliferation occurred during the high-temperature growth process, and the pseudomorphic growth states were disrupted. Conversely, the RSM pattern of the segmented temperature sample exhibits sharp and concentrated diffraction spots without broadening or shifting, indicating a fully coherent strain state.

Furthermore, the surface roughness of the samples was characterized by AFM (**Figure 4**). The two constant-temperature samples exhibit higher surface roughness and a slight tendency toward three-dimensional island growth. In contrast, the segmented temperature sample shows a flat surface with a root mean square roughness (Rq) of only 0.08 nm over a 10 × 10 $\mu m^2$ area, far below 1 nm, maintaining a perfect two-dimensional growth mode.

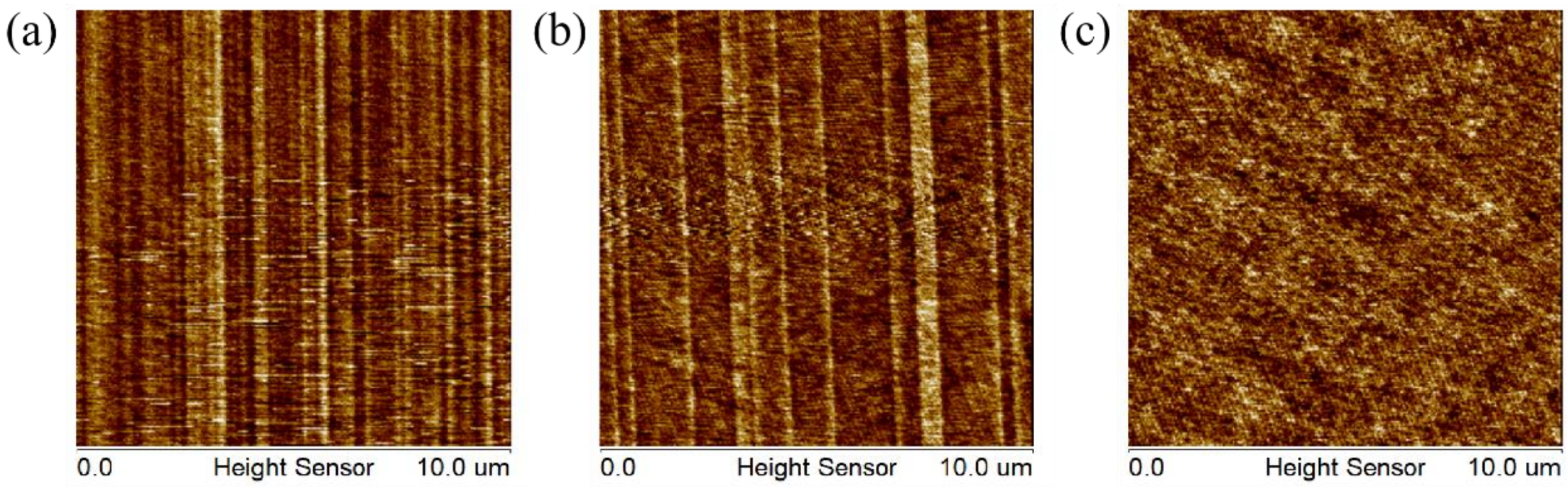

**Figure 4.** 10 × 10 $\mu m^2$ AFM images of the 4 + 4 channel superlattice structures: (a) grown under 600 °C, (b) grown under 650 °C, and (c) grown under segmented 650–600 °C.

This stark contrast indicates that a fixed single temperature growth mode cannot overcome the thermal budget accumulation effect in multi-layer stacks. The segmented temperature control strategy, which effectively suppresses diffusion and increases the critical thickness through lower-temperature growth in the later stages, is an essential approach for fabricating high-quality 4 + 4 channel structures.

## Quality of Multi-Channel Superlattice Structures under Segmented Temperature Control

After validating the effectiveness of the segmented temperature control strategy, we applied it to 2 + 2, 3 + 3, and 4 + 4 channel superlattice structures to investigate its scalability and the evolution of crystal quality.

As shown in **Figure 5(a)**, the ω-2θ HRXRD scans around the Si (004) Bragg reflection for the 2 + 2, 3 + 3, and 4 + 4 channel superlattice structures, while **Figures 5(b) – (d)** show the corresponding (113) RSM results. The data show that increasing the number of channels from two to three leads to broadening and intensity reduction of the superlattice satellite peaks. This change reflects increased strain accumulation and greater difficulty in interface control in multi-channel stacks. This readily induces misfit dislocation proliferation and strain relaxation, which can disrupt the periodicity and crystal integrity of the superlattice. Notably, the 4 + 4 channel superlattice structure maintains clear higher-order satellite peaks and a favorable strained growth state, even under the stringent conditions of further increased channel count and a narrowed epitaxial growth window. The diffraction spots in the RSM remain well-concentrated, with no significant lateral broadening or dispersion. This improvement is attributed to the segmented temperature control strategy. Lowering the growth temperature during the mid-to-late growth stages, the relaxation tendency of the middle and bottom SiGe layers under prolonged high-temperature exposure is significantly suppressed. This results in markedly reduced interdiffusion at the bottom interfaces and allows the top layers to maintain perfect pseudomorphic growth at the lower temperature. This 4 + 4 channel structure substantially increases the number of channels to enhance the current drive capability of the CFET device while fully ensuring the overall crystal quality.

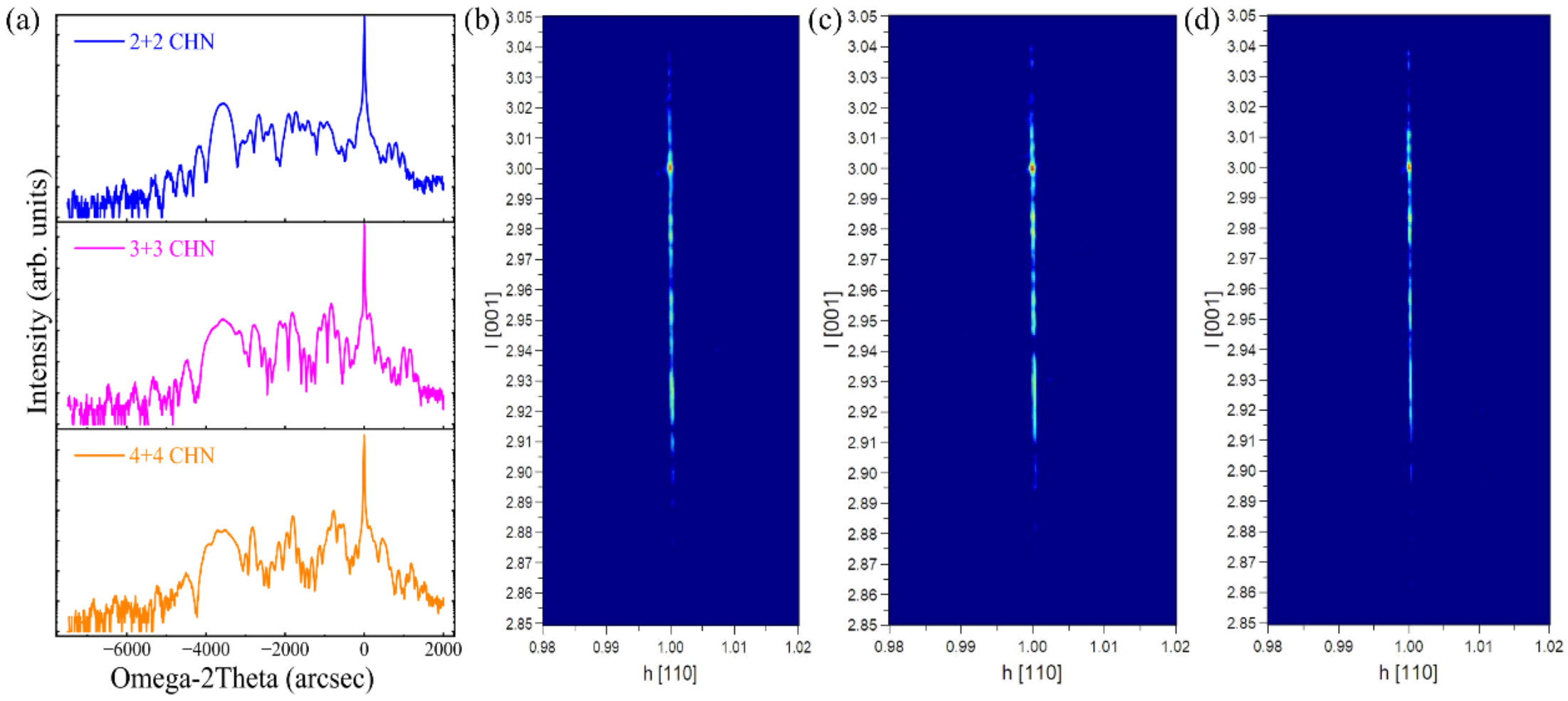

**Figure 5.** (a) XRD ω-2θ scan around the Si (004) Bragg reflection for the 2 + 2, 3 + 3, and 4 + 4 channel superlattice structures for the CFET stacks, and (b) – (d) corresponding RSM around the Si (113).

As the stacking period expands from 2 + 2 channels to 4 + 4 channels, the thermal budget experienced by the bottom epitaxial layers rises significantly, leading to aggravated lattice strain accumulation and exponentially increasing the risk of Si-Ge interdiffusion. To thoroughly investigate the structural integrity and interface thickness under multi-layer stacking, HRTEM and EDS were employed for atomic-level characterization of the superlattice cross sections.

**Figure 6** presents the TEM interface images and corresponding EDS line scans for the 2 + 2, 3 + 3 and 4 + 4 channel superlattice structures. All structures exhibit excellent pseudomorphic growth mode. In the TEM cross-sectional micrographs images, all samples exhibit straight and well-defined interfaces without significant interface roughness or layer thickness fluctuations. No epitaxial defects such as misfit dislocations or stacking faults are observed throughout the entire stacked region, indicating high crystal quality. The EDS line scans indicate sharp interfaces with narrow transition regions and negligible interdiffusion for the 2 + 2, 3 + 3, and 4 + 4 channel structures. Specifically, for the 4 + 4 channel superlattice structure, the low-Ge $Si_{0.8}Ge_{0.2}$ layers maintain a Ge concentration of approximately 20 %, while the high-Ge $Si_{0.6}Ge_{0.4}$ layers sustain a Ge concentration around 39 %. The average thicknesses of the Si and $Si_{0.8}Ge_{0.2}$ layers are approximately 10 nm and 8.97 nm, with standard deviations of 1.19 nm and 0.26 nm, respectively. The high-Ge $Si_{0.6}Ge_{0.4}$ layers measure 30.75 nm in thickness. These results demonstrate that the layer thicknesses and concentrations of the 4 + 4 channel superlattice structure are in good agreement with the design targets, exhibiting excellent thickness and compositional uniformity.

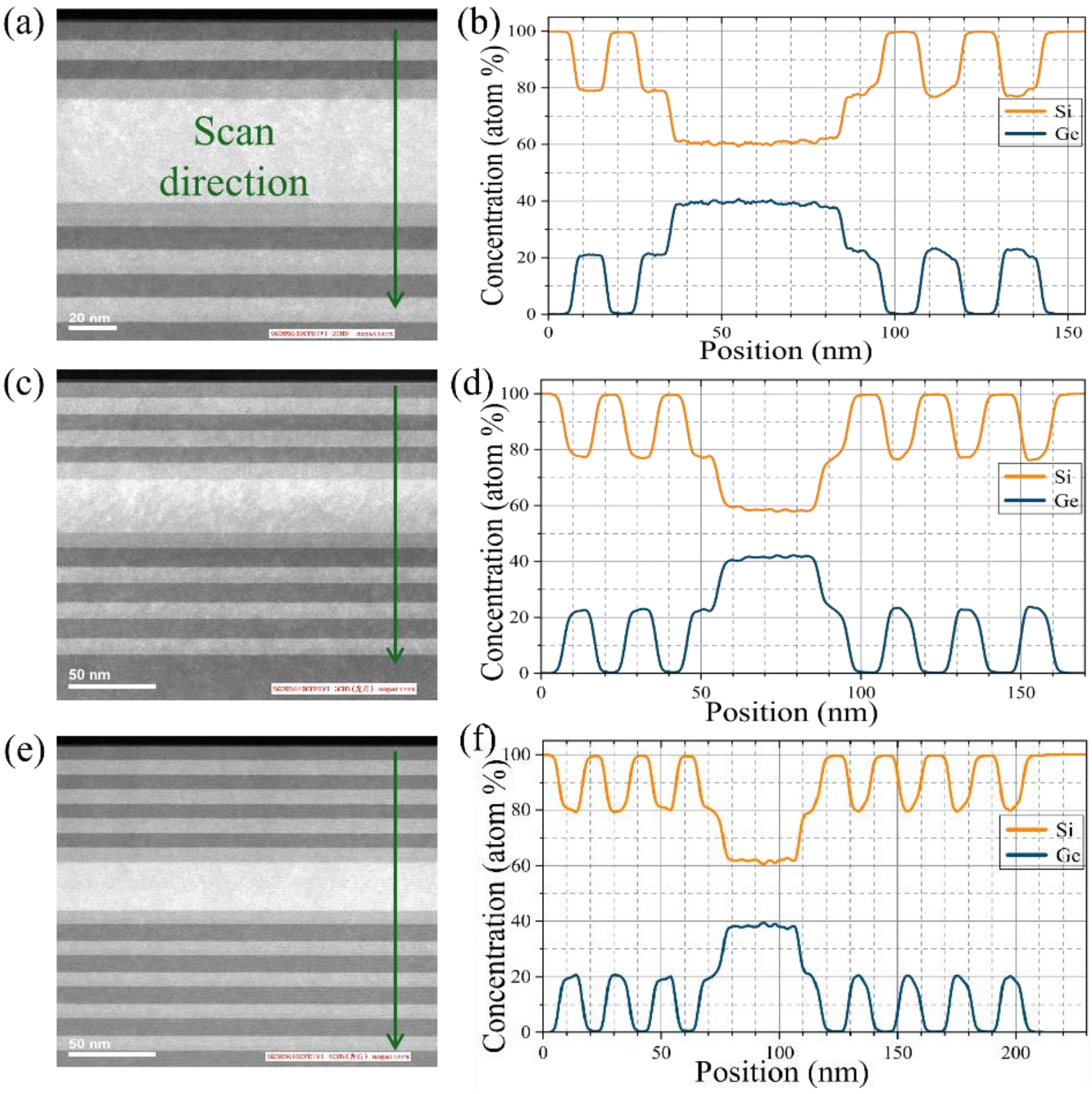

**Figure 6**. Cross-sectional TEM images and corresponding EDS line scans of the (a, b) 2 + 2 channel, (c, d) 3 + 3 channel, and (e, f) 4 + 4 channel Si/SiGe superlattice structures.

In addition to the direct characterization of interface abruptness, we performed quantitative analysis to further validate the results. HAADF-STEM contrast imaging analysis was conducted on the interfaces of the 4 + 4 channel sample. Since the contrast is proportional to the square of the atomic number, this method allows extraction of the Ge contrast distribution. The procedure involved integrating the TEM image to extract the contrast, followed by fitting with a sigmoid function expressed as:

$$f(x) = \frac{C}{1 + e^{\frac{\pm(x_0 - x)}{\tau}}} + V \tag{2}$$

where C and V are constants, $x_0$ represents the interface position, and 4τ is used to denote the interface width. It should be noted that the extracted $4\tau$ values inherently include the effects of interface roughness, resulting in measured thicknesses larger than the actual physical interface, corresponding approximately to the region where the normalized Ge content transitions from 14% to 86% of its maximum [20, 30]. **Figure 7(a)** exemplarily shows a partial STEM image of the 4 + 4 channel sample and the corresponding Ge contrast distribution profile extracted from it. And the Ge contrast distribution curves obtained after fitting are presented in **Figures 7(b) – (d)**.

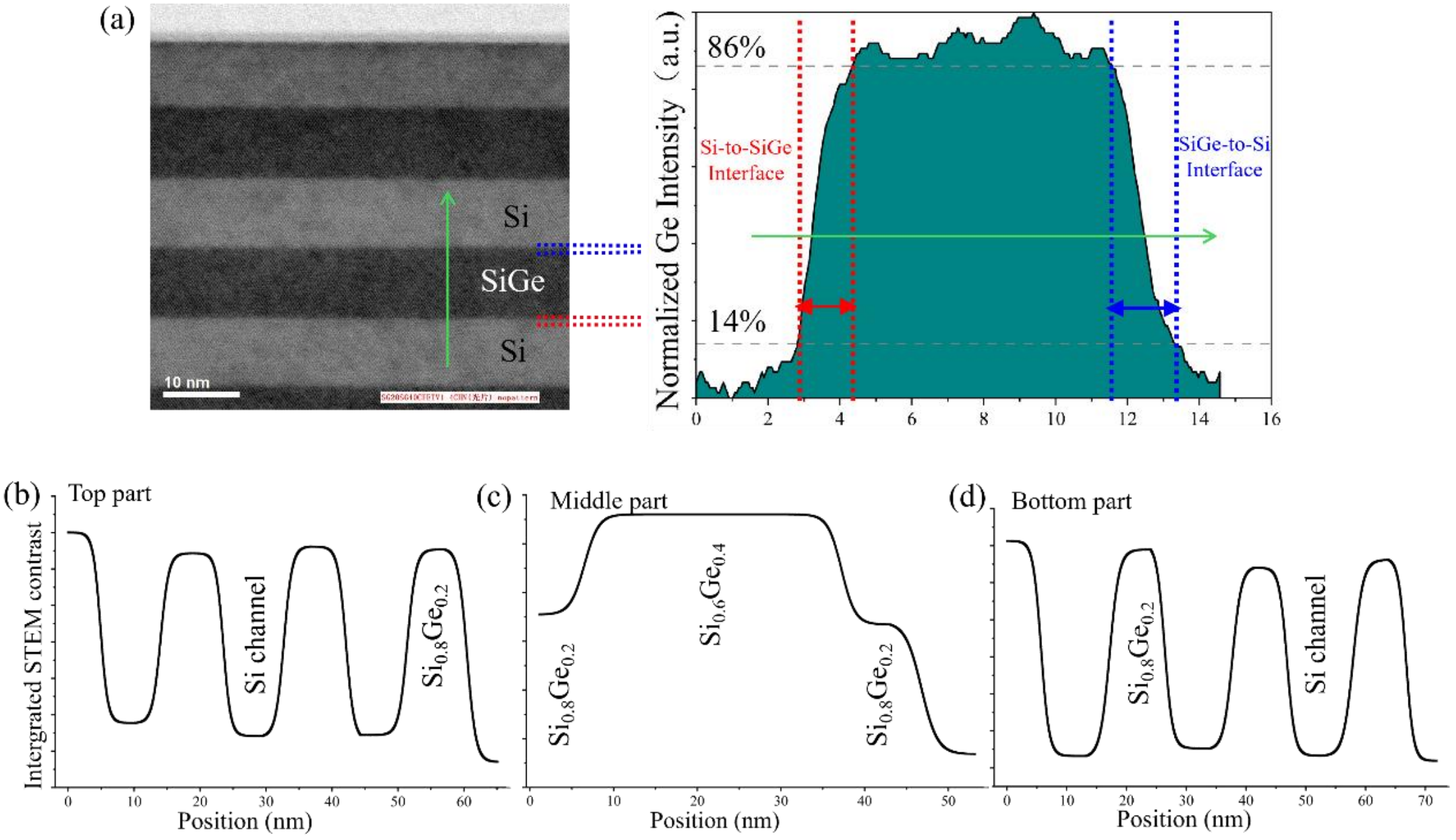

**Figure 7.** (a) Cross-sectional HAADF-STEM (Z-contrast) image of the four-channel (4 + 4) Si/SiGe superlattice structure for CFET devices. The red (blue) dashed lines indicate the Si-to-SiGe (SiGe-to-Si) interfaces, and the red (blue) double-headed arrows mark the transition width, defined as the region where the Ge intensity changes from 14% to 86% of its maximum. (b–d) Horizontally integrated contrast profiles extracted from the high-resolution HAADF-STEM images.

**Figure 8** summarizes the Ge concentration and corresponding interface thickness (4τ) of each layer from the bottom (B1–B4) to the top (T1–T4) of the 4 + 4 channel sample, extracted from EDS and the Ge contrast profiles. As shown in **Figure 8(a)**, the average Ge concentration in the bottom $Si_{0.8}Ge_{0.2}$ layers is approximately 18.5 %, significantly lower than the 19.5 % observed in the top layers. This difference is attributed to the bottom stacks experiencing a longer high-temperature exposure throughout epitaxial growth, resulting in a higher accumulated thermal budget. This drives Ge atoms to diffuse outward from the SiGe layers into adjacent layers, thereby diluting the effective Ge concentration within the SiGe layers. Correspondingly, the interface thickness data in **Figure 8(b)** further support this mechanism: both Si-to-SiGe and SiGe-to-Si interface thicknesses in the bottom region are substantially thicker than those in the top region, directly confirming that thermal budget-induced atomic interdiffusion is the primary cause of interface and composition degradation.

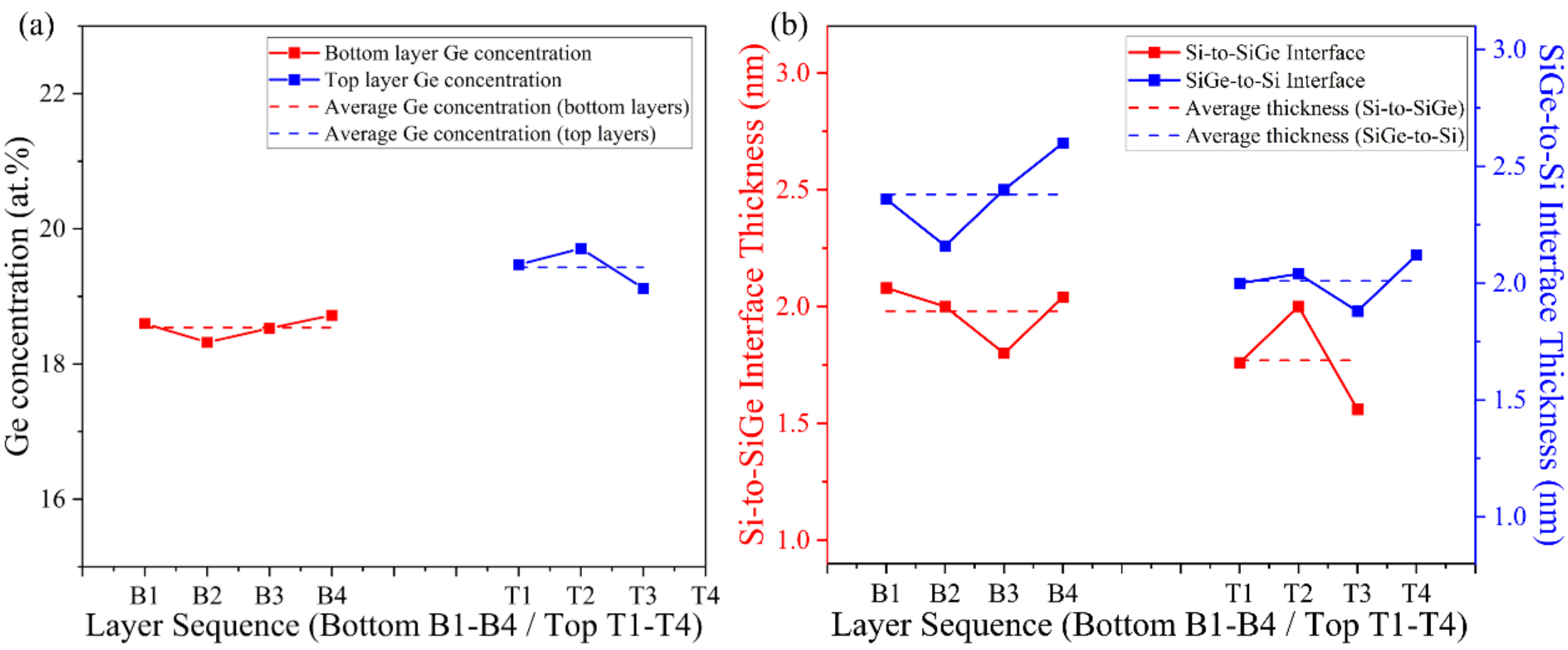

**Figure 8. (a)** Si-to-SiGe interface thickness and (b) average Ge concentration as a function of layer position from bottom (B1–B4) to top (T1–T4) in the 4 + 4 channel superlattice structure. The red and blue dashed lines indicate the average Ge concentration for bottom and top layers, respectively.

Analysis of different interface types indicates that SiGe-to-Si interfaces are consistently thicker than Si-to-SiGe interfaces across both bottom and top stack. This is commonly attributed to Ge segregation occurring when Si is grown on a SiGe surface. The segmented temperature control strategy adopted in this study significantly suppresses Ge diffusion and segregation by lowering the growth temperature during the top stack deposition, thereby greatly improving interface abruptness.

To further demonstrate the progress of this work, the key structural parameters of the Si/SiGe superlattice stack developed in this study are summarized in **Table 1**, together with the literature data reported by Huang et al. [15]and Loo et al. [16].

All three superlattice structures are fully coherent and exhibit low defect densities, indicating high crystal quality. Both this work and Huang et al. achieve a higher channel count (4 + 4 vs. 3 + 3), thereby enhancing channel integration density, and also obtain higher Ge concentrations in both the low-Ge layer (20 % vs. 17.7 %) and the high-Ge layer (40 % vs. 37 %). The increased Ge concentration enhances strain, which can significantly improve selective etching ratios and positively benefits device performance while broadening the process window. In terms of interface thickness control, this work achieves excellent interface abruptness through the segmented temperature control strategy. The interface thickness of the 4+4 channel structure in this work is comparable to the 1 + 1 channel structure reported by Loo et al. [16]. In addition, the Si channel layers in this study are thicker than those in prior reports (9 nm in Loo et al. [16] and approximately 6 nm in Huang et al. [15]), which can enhance device current-driving capability and overall performance. Importantly, this work quantifies, for the first time, the evolution of interface thickness and Ge concentration from the bottom to the top of the stack. The segmented temperature control strategy (650–600 °C) leads to sharper top interfaces and a composition closer to the design target, highlighting its advantage in suppressing interdiffusion in high-stack structures.

In summary, the experimental results demonstrate that the bottom stacks, due to prolonged high-temperature exposure, exhibit significant interface broadening and Ge concentration deviation from the target value, whereas the top stack, protected by the segmented temperature control strategy, maintain sharp interfaces and uniform compositions. This directly proves that thermal budget accumulation is the primary cause of interface degradation in multi-channel superlattices, while also validating the necessity and effectiveness of the proposed segmented temperature control strategy in suppressing Ge interdiffusion, maintaining interface abruptness, and preserving composition accuracy. Thanks to this strategy, the 4 + 4 channel superlattice structure maintains excellent crystal quality and interface characteristics even under a total thickness exceeding 210 nm, providing a high-quality material foundation for the precise release of nanosheets in CFET devices.

**Table 1.** Comparison of Si/SiGe superlattice structures reported in this work and in the literature (Huang et al. [15] and Loo et al. [16]) for CFET applications.

| Parameter | This work | Loo et al. [16] | Huang et al. [15] |
| --- | --- | --- | --- |
| Si channel count | 4 + 4 channel | 3 + 3 channel | 4 + 4 channel |
| Si liner | No | Yes | No |
| Low Ge concentration | $Si_{0.8}Ge_{0.2}$ | $Si_{0.823}Ge_{0.177}$ | $Si_{0.8}Ge_{0.2}$ |
| High Ge concentration | $Si_{0.6}Ge_{0.4}$ | $Si_{0.63}Ge_{0.37}$ | $Si_{0.6}Ge_{0.4}$ |
| Growth method | Segmented 650–600 °C | Relatively high temperature | — |
| Si-to-SiGe interface thickness | 1.5–2 nm | 1.52 nm (1 + 1 channel top part)<br>1.58 nm (1 + 1 channel bottom part) | — |
| SiGe-to-Si interface thickness | 1.8–2.6 nm | 2.1 nm (1 + 1 channel top part)<br>1.89 nm (1 + 1 channel bottom part) | — |
| Layer thickness | Si channel (10 nm)<br>$Si_{0.8}Ge_{0.2}$ (9 nm)<br>Total (210 nm) | Si channel (9 nm)<br>$Si_{0.823}Ge_{0.177}$ (9 nm)<br>Total (185 nm) | Si channel (6 nm)<br>$Si_{0.8}Ge_{0.2}$ (9 nm)<br>Total (160 nm) |

**4. Conclusion**

This study proposes and validates a segmented temperature control strategy for epitaxial growth, successfully fabricating a high-quality 4 + 4 channel Si/SiGe superlattice structure for advanced logic device applications. By reducing the growth temperature to 600 °C in the later stages, the interdiffusion coefficient of Ge atoms is suppressed to approximately 5.6–7% of its original value, effectively mitigating interfacial interdiffusion and strain relaxation in the bottom stacks caused by prolonged high-temperature exposure. Consequently, the 4 + 4 channel superlattice, with a total thickness exceeding 210 nm, maintains a fully coherent strain state, atomically abrupt interface thickness, and excellent crystalline integrity. Compared with the existing literature, this work achieves a higher channel count (4 + 4), higher Ge content (40 %), and a thicker Si channel layer (10 nm), which are beneficial for improving device performance and broadening the process window. Furthermore, this work provides the first quantitative analysis of the evolution of interface thickness and Ge concentration from the bottom to the top of the stack, directly confirming that thermal budget accumulation is the root cause of interface degradation in multi-channel superlattices. This study not only deepens the understanding of interdiffusion mechanisms in Si/SiGe superlattices but also provides a reliable process pathway and theoretical basis for the epitaxial integration of high-channel-count logic devices.

**Declaration of competing interest**

The authors declare that they have no known competing financial interests or personal relationships that could have appeared to influence the work reported in this paper.

**Acknowledgments**

This work was supported by the Beijing Natural Science Foundation (Grant No. Z220006), Joint Laboratory Project between the Institute of Microelectronics, Chinese Academy of Sciences and Beijing NAURA Microelectronics Equipment Co., Ltd (Grant No. NMC-CG20243972FE).

**Data availability statements**

All data generated or analyzed during this study are included in this published article [and its supplementary information files].